\documentclass[review,12pt]{elsarticle}

\usepackage{amssymb}
\usepackage{amsthm}
\usepackage{amsmath}

\usepackage{mathrsfs}
\usepackage{graphicx}
\usepackage{epstopdf}
\usepackage{float}
\usepackage{caption}
\usepackage{subcaption}
\usepackage{bm}
\usepackage{bbm}
\usepackage{mathrsfs}
\usepackage{cleveref}
\usepackage{soul}
\usepackage{accents}
\usepackage{graphicx}
\usepackage{xcolor}
\usepackage{courier} 
\usepackage{listings} 
\usepackage[T1]{fontenc}
\usepackage{lmodern}
\usepackage{tabu} 
\usepackage{longtable}
\usepackage{changepage} 
\usepackage[margin=2.2cm]{geometry}
\biboptions{numbers,sort&compress}

\usepackage{booktabs}
\usepackage{array, multirow}
\usepackage{microtype}

\journal{Applied Energy}

\makeatletter
\def\@author#1{\g@addto@macro\elsauthors{\normalsize%
    \def\baselinestretch{1}%
    \upshape\authorsep#1\unskip\textsuperscript{%
      \ifx\@fnmark\@empty\else\unskip\sep\@fnmark\let\sep=,\fi
      \ifx\@corref\@empty\else\unskip\sep\@corref\let\sep=,\fi
      }%
    \def\authorsep{\unskip,\space}%
    \global\let\@fnmark\@empty
    \global\let\@corref\@empty  
    \global\let\sep\@empty}%
    \@eadauthor={#1}
}
\makeatother

\begin{document}

\begin{frontmatter}



\title{Phase field modelling of cracking and capacity fade in core-shell cathode particles for lithium-ion batteries}


\author[IC,Oxf]{Yang Tu}

\author[IC1,FI]{Billy Wu}

\author[Oxf,IC]{Emilio Mart\'{\i}nez-Pa\~neda\corref{cor1}}
\ead{emilio.martinez-paneda@eng.ox.ac.uk}

\address[IC]{Department of Civil and Environmental Engineering, Imperial College London, London SW7 2AZ, UK}
\address[IC1]{Dyson School of Design Engineering, Imperial College London, London SW7 2AZ, UK}
\address[FI]{The Faraday Institution, Quad One, Becquerel Avenue, Harwell Campus, Didcot, OX11 0RA, UK}
\address[Oxf]{Department of Engineering Science, University of Oxford, Oxford OX1 3PJ, UK}

\cortext[cor1]{Corresponding author.}

\begin{abstract} 
Core-shell electrode particles are a promising morphology control strategy for high-performance lithium-ion batteries. However, experimental observations reveal that these structures remain prone to mechanical failure, with shell fractures and core-shell debonding occurring after a single charge. In this work, we present a novel, comprehensive computational framework to predict and gain insight into the failure of core-shell morphologies and the associated degradation in battery performance. The fully coupled chemo-mechano-damage model presented captures the interplay between mechanical damage and electrochemical behaviours, enabling the quantification of particle cracking and capacity fade. Both bulk material fracture and interface debonding are captured by utilising the phase field method. We quantify the severity of particle cracking and capacity loss through case studies on a representative core-shell system (NMC811@NMC532). The results bring valuable insights into cracking patterns, underlying mechanisms, and their impact on capacity loss. Surface cracks are found to initiate when a significantly higher lithium concentration accumulates in the core compared to the shell. Interfacial debonding is shown to arise from localised hoop stresses near the core-shell interface, due to greater shell expansion. This debonding develops rapidly, impedes lithium-ion transport, and can lead to more than 10\% capacity loss after a single discharge. Furthermore, larger particles may experience crack branching driven by extensive tensile zones, potentially fragmenting the entire particle. The framework developed can not only bring new insight into the degradation mechanisms of core-shell particles but also be used to design electrode materials with improved performance and extended lifetime.\\
\end{abstract}



\end{frontmatter}


\section{Introduction}
\label{Introduction}
Electrode particle cracking is one of the main causes of lithium-ion battery degradation \cite{li2018single,Edge2021}. This leads to loss of active material \cite{friedrich2019capacity} and increases the exposed surface area, accelerating the growth of solid–electrolyte and cathode-electrolyte interphases, resulting in capacity fade and impedance rise. Particle cracking often arises from mechanical stress induced by expansion and shrinkage of the active material during lithium intercalation and deintercalation \cite{su2021multiscale}, with pre-existing cracks commonly introduced during the fabrication process, particularly during calendering \cite{zhang2024deformation,Parks2023}. To mitigate particle cracking, several microstructure-control approaches have been proposed, including core-shell \cite{Chen2010,zhong2020nano,chen2017recent,li2022b}, concentration-gradient \cite{Sun2009,sun2010novel,ju2014optimization}, single-crystal \cite{li2017comparison,li2018synthesis,liu2020microstructural}, and radially aligned \cite{xu2019radially,kim2019microstructure} particles. In this work, we focus on the core-shell particle architecture, which is a popular and cost-effective approach. Core-shell particles typically consist of a high specific capacity but chemically reactive material core with a more stable, yet lower specific capacity shell. This structure reduces the exposure of the core to the electrolyte, creating a more stable chemical environment. Moreover, the shell can restrict active material volume changes, thereby minimising cracking. However, experiments have shown that core-shell structures are still susceptible to mechanical failure, with shell fracture and core-shell debonding being observed after just a single charge at low currents (\textit{e.g.} C/3) \cite{Brandt2020}.\\

To better understand mechanical degradation, research efforts have examined particle cracking mechanisms. Advanced imaging techniques, including X-ray computed tomography (XCT) and scanning electron microscopy (SEM), have been employed to visualise cracking patterns in electrode particles  \cite{Parks2023,Lin2021}. 
Continuum modelling methods have been used to investigate diffusion-induced stress profiles and predict crack growth. 
Furthermore, advanced modelling approaches such as the phase field method, which is a variational formulation of Griffith's fracture theory \cite{bourdin2000numerical,kristensen2021assessment}, enable the prediction of complex cracking patterns in arbitrary geometries. For instance, Miehe \textit{et al.} \cite{Miehe2016} proposed a phase field modelling framework for chemo-mechanical induced fracture in both 2D and 3D electrode particles systems. Similarly, Klinsmann \textit{et al.} \cite{klinsmann2015modeling,klinsmann2016modeling} investigated the effects of particle size, initial crack size, and charging rate on the cracking behaviours of $\mathrm{LiMn_2O_4}$ particles. Boyce and co-workers \cite{Boyce2022,boyce2024role} combined chemo-mechanical phase field fracture modelling with XCT imaging, gaining insight at both the particle and electrode levels. Ai \textit{et al.} \cite{Ai2022} incorporated fatigue degradation into the phase field model to predict particle cracking. Additionally, they have applied the model to 3D particle geometries scanned by XCT, providing insights into the cracking behaviours of realistic microstructures. 
However, the application of phase field method to core-shell particles remains limited \cite{roque2024phase,shen2024core}. Moreover, the electrochemical consequences of particle cracking, such as capacity fade and impedance rise \cite{han2024computational,Allen2021}, have not been addressed yet in the context of core-shell structures, despite their widespread adoption in state-of-the-art cathode design. This is an area where modelling insight is strongly needed due to the challenges associated with the simultaneous experimental assessment of mechanical degradation and electrochemical performance at the particle level.\\

In this work, we propose a fully coupled chemo-mechano-damage framework to unravel cracking and capacity fade in core-shell electrode particles. We establish a link between particle cracking and capacity degradation, offering a new approach to understanding electrochemical degradation at the particle level. Bulk material fracture and interfacial debonding are simultaneously addressed by using a novel, phase field-based numerical strategy. Model predictions are validated against experimental observations of particle cracking. Subsequently, case studies on a high-nickel core–shell system are conducted to investigate cracking mechanisms and assess the influence of key design and operating parameters. The findings provide valuable guidance for designing more durable electrode particles to mitigate mechanical damage and minimise capacity degradation.

\section{Methods}
\label{sec:Methods}
\begin{figure}[H]
    \centering
    \includegraphics[width=0.8\textwidth]{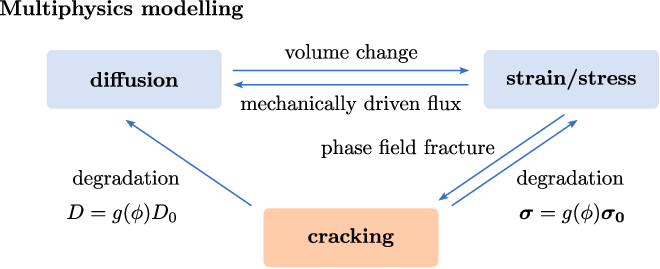}
    \caption{Schematic representation of the multiphysics framework, showing the coupling among lithium-ion diffusion, strain/stress, and cracking.}
    \label{fig:coupling}
\end{figure}

We proceed to describe our theory, which includes the coupling between lithium-ion diffusion and mechanical stress (Section \ref{subsec:diffusion}), the description of phase field fracture (Section \ref{subsec:pf}), a diffuse representation of the interface to model debonding (Section \ref{subsec:diffuse}), and the interplay between mechanical damage and diffusion (Section \ref{subsec:coupling}). Finally, we outline the boundary conditions used for the intercalation process. The multiphysics framework is schematically illustrated in Fig. \ref{fig:coupling}.

\subsection{Diffusion-induced stress: coupled model of diffusion and mechanics}
\label{subsec:diffusion}
The conservation of species balance gives
\begin{equation}
    \frac{\partial c}{\partial t}+\bm{\nabla\cdot \mathrm{J}}=0, 
\end{equation}
\noindent where $t$ represents time, $c$ denotes the lithium-ion concentration in the active material, and $\bm{\mathrm{J}}$ is the flux of lithium ions. The chemical potential gradient drives the movement of lithium ions, with the flux being proportional to the gradient of chemical potential \cite{zhang2007numerical}:
\begin{equation}
    \bm{\mathrm{J}}=-Mc\bm{\nabla} \mu,
    \label{eq:flux}
\end{equation}
where $M=D/RT$ is the mobility of lithium ions in the host material, and $\mu$ is the chemical potential. Based on thermodynamics, the chemical potential in an ideal solid solution can be expressed as \cite{wang2002effect,zhang2007numerical}:
\begin{equation}
    \mu=\mu_0+RT\ln(c)-\Omega\sigma_h,
    \label{eq:mu1}
\end{equation}
where $\mu_0$ is the reference chemical potential, \(R\) is the gas constant, \(T\) is the temperature, \(\Omega\) is the partial molar volume of lithium ions in the host material, and \(\sigma_h=\mathrm{tr}(\bm{\sigma})/3\) is the hydrostatic stress, with $\bm{\sigma}$ being the Cauchy stress tensor. Combining Eqs. (\ref{eq:flux}) and (\ref{eq:mu1}), the species flux can be expressed as 
\begin{equation}
    \bm{\mathrm{J}}=-D\bm{\nabla} c+\frac{cD\Omega}{RT}\bm{\nabla}\sigma_h\quad \mathrm{and}\quad \bm{\mathrm{J}}\cdot\bm{\mathrm{n}}=J\quad \mathrm{at\ } \partial V_s,
    \label{eq:J}
\end{equation}
where \(D\) is the diffusion coefficient, \(J\) is the flux magnitude at the boundary, and $V_s$ denotes the domain of the solid. The first term in the right-hand side of Eq. (\ref{eq:J}) captures the concentration gradient contribution to the flux, while the second term captures the role of mechanical field (lattice distortion) in driving Li transport. Conversely, the concentration field drives volume changes, thereby establishing a two-way coupling between diffusion and mechanics. The constitutive relation between the Cauchy stress tensor \(\bm{\sigma_0}\) of an undamaged solid and the strain tensor \(\bm{\varepsilon}\) is given by:
\begin{equation}
\label{eq:ConstModel}
    \bm{\sigma_0}=\lambda\mathrm{tr}(\bm{\varepsilon-\varepsilon_{Li}})\bm{I}+2 G (\bm{\varepsilon-\varepsilon_{Li}}),
\end{equation}
where \(\lambda\) and \(G\) are the Lamé constants, and $\bm{I}$ is the identity matrix. In Eq. (\ref{eq:ConstModel}), \(\bm{\varepsilon_{Li}}\) is the chemical strain caused by lithium-ion insertion which can be expressed as 
\begin{equation}
    \bm{\varepsilon_{Li}}=\frac{1}{3}\Omega\left(c-c_0\right)\bm{I},
\end{equation}
where \(c_0\) represents the initial lithium-ion concentration in a stress-free state.

\subsection{Phase field fracture formulation}
\label{subsec:pf}
\begin{figure}[H]
    \centering
    \includegraphics[width=1\textwidth]{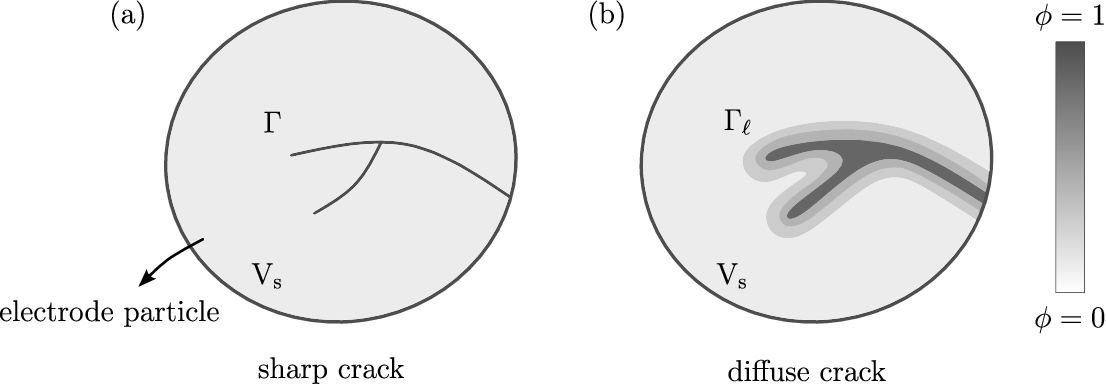}
    \caption{Schematic representation of an electrode particle of volume $V_s$ containing (a) a discrete sharp crack, described by $\Gamma$ and (b) a phase field crack, characterised by the phase field order parameter $\phi$ and the regularised crack surface $\Gamma_\ell$.}
    \label{fig:crack}
\end{figure}

According to Griffith's theory \cite{griffith1921vi}, a crack will propagate when the potential energy released due to crack growth equals or exceeds the energy needed to create new free surfaces. For a solid with strain energy density $\psi(\boldsymbol{\varepsilon})$ which is a function of the strain tensor $\boldsymbol{\varepsilon}$, the variation in total energy $\Pi$ resulting from an incremental change in the crack area $\mathrm{d} A$, in the absence of external forces, is described by:
\begin{equation}
\frac{\mathrm{d} \Pi}{\mathrm{d} A}=\frac{\mathrm{d} \psi(\boldsymbol{\varepsilon})}{\mathrm{d} A}+\frac{\mathrm{d} W_c}{\mathrm{d} A}=0,
\end{equation}
where $W_c$ is the work needed to create new crack surfaces. The last term, $\mathrm{d} W_c / \mathrm{d} A = G_c$, represents a material property that characterises the material toughness, commonly known as the critical energy release rate or material toughness. In a variational form, Griffith's energy balance is given by \cite{francfort1998revisiting}:
\begin{equation}
\Pi=\int_{V_s} \psi(\boldsymbol{\varepsilon)}\ \mathrm{d} V+\int_{\Gamma} G_c \ \mathrm{d} \Gamma,
\label{eq:Pi}
\end{equation}
with $\Gamma$ being the crack surface and $V_s$ being the domain of the solid. Predicting crack growth by minimising Eq. (\ref{eq:Pi}) is challenging due to the unknown nature of $\Gamma$. The variational phase field method offers a promising computational approach to address this numerical difficulty. A scalar auxiliary variable, the phase field $\phi$, is introduced here to describe cracks using a diffuse representation rather than a discrete discontinuity, as illustrated in Fig. \ref{fig:crack}. The phase field $\phi$ describes the degree of damage sustained by the material, with this varying between 0 (intact) and 1 (fully cracked). Accordingly, the Griffith functional (\ref{eq:Pi}) can be approximated as the following:
\begin{equation}
    \Pi_{\ell}=\int_{V_s}\left[ \psi(\boldsymbol{\varepsilon},\phi)+G_c \gamma(\phi, \ell)\right] \mathrm{d} V,
    \label{eq:gamma}
\end{equation}
where $\ell$ is a length scale parameter that governs the width of the diffuse crack zone, and $\gamma$ is the crack surface density function that depends on $\ell$ and $\phi$. Eq. (\ref{eq:gamma}) is expressed as a volume integral and can be solved computationally, without the need to know the crack surface $\Gamma$ \textit{a priori}. Following the work by Bourdin \textit{et al.} \cite{bourdin2000numerical}, which was inspired by Ambrosio and Tortorelli \cite{ambrosio1990approximation}, we employ the so-called AT2 model and define $\gamma$ as:
\begin{equation}
    \gamma(\phi, \ell)=\frac{1}{2 \ell} \phi^2+\frac{\ell}{2}|\nabla \phi|^2,
\end{equation}

The strain energy density $\psi$ is degraded with increasing damage. To achieve this, we employ a degradation function $g(\phi)$. Since we only consider cracking induced by tensile stresses, the degradation function is applied solely to the tensile part of the strain energy:
\begin{equation}
\psi(\boldsymbol{\varepsilon},\phi)=g(\phi)\psi_0^+(\boldsymbol{\varepsilon})+\psi_0^-(\boldsymbol{\varepsilon}),
\end{equation}
where $\psi_0^+$ and $\psi_0^-$ are respectively the (undamaged) tensile and compressive strain energy densities, which are defined as follows based on 
the volumetric-deviatoric split proposed by Amor \textit{et al.} \cite{amor2009regularized}: 
\begin{align}
& \psi_0^{+}=0.5 K\left\langle\operatorname{tr}\left(\boldsymbol{\varepsilon}-\boldsymbol{\varepsilon_{\mathrm{Li}}}\right)\right\rangle_{+}^2+\mu\left(\boldsymbol{\varepsilon^{\mathrm{dev}}}: \boldsymbol{\varepsilon^{\mathrm{dev}}}\right), \\
& \psi_0^{-}=0.5 K\left\langle\operatorname{tr}\left(\boldsymbol{\varepsilon}-\boldsymbol{\varepsilon_{\mathrm{Li}}}\right)\right\rangle_{-}^2,
\end{align}
where $K=\lambda+2G/3$ is the bulk modulus, $\langle x\rangle_{ \pm}=(x \pm|x|) / 2$ are the Macaulay brackets, and $\boldsymbol{\varepsilon}^{d e v}=\left(\boldsymbol{\varepsilon}-\boldsymbol{\varepsilon_{\mathrm{Li}}}\right)-\operatorname{tr}\left(\boldsymbol{\varepsilon}-\boldsymbol{\varepsilon_{\mathrm{Li}}}\right) \boldsymbol{I} / 3$ is the deviatoric elastic strain. The degradation function $g(\phi)$ must monotonically decrease and satisfy the conditions $g(0)=1,\ g(1)=0$, and $g^{\prime}(1)=0$. The widely used quadratic form is adopted here: 
\begin{equation}
    g(\phi)=(1-\phi)^2 \, .
\end{equation}
Thus, the general form of the Griffith functional (\ref{eq:gamma}) becomes
\begin{equation}
    \Pi_{\ell}=\int_{V_s}\left[ (1-\phi)^2\psi_0^+(\boldsymbol{\varepsilon})+\psi_0^-(\boldsymbol{\varepsilon})+G_c \left(\frac{1}{2 \ell} \phi^2+\frac{\ell}{2}|\nabla \phi|^2\right)\right] \mathrm{d} V,
\end{equation}
For numerical purposes, we use $g(\phi)=(1-\phi)^2+k$ where $k$ is a small positive constant to avoid numerical issues when $\phi=1$. Next, we introduce a history variable field, $\mathcal{H}$, to ensure irreversible growth of the phase ﬁeld variable. The history variable is defined as $\mathcal{H}=\max _{\tau \in[0, t]} \psi_0^+(\tau)$, which tracks the maximum value of the tensile strain energy over time \cite{ambati2015review}. Finally, based on the constitutive formulation of the dissipation and the energy storage in a cracked solid, we can derive the strong form balance equations using the principle of virtual power. For a quasi-static process, the governing equations include the static equilibrium condition for the damaged stress $\sigma$ and the evolution equation for the phase field:
\begin{align}
\boldsymbol{\nabla} \cdot\boldsymbol{\sigma}=\mathbf{0}\quad \mathrm{with}&\quad \boldsymbol{\sigma}=g(\phi) \boldsymbol{\sigma_0}, \\
\frac{G_{c}}{\ell}\left(\phi-\ell^2 \nabla^2 \phi\right)&=2(1-\phi) \mathcal{H},
\end{align}
The boundary conditions are given as
\begin{align}
& \boldsymbol{\sigma} \cdot \mathbf{n}=\overline{\mathbf{t}} \quad \text { and } \quad \mathbf{u}=\overline{\mathbf{u}} \quad \text { at } \partial V_s, \\
& \boldsymbol{\nabla} \phi \cdot \mathbf{n}=0 \quad \text { at } \partial V_s,
\end{align}
where $\partial V_s$ is the boundary of the solid, $\mathbf{n}$ is the outward normal, $\overline{\mathbf{t}}$
is the external force vector, $\mathbf{u}$ is the
displacement vector, and $\overline{\mathbf{u}}$ denotes the displacement constraint.

\subsection{Diffuse zone of core-shell bonding}
\label{subsec:diffuse}
The phase field fracture model is also capable of capturing debonding \cite{au2023hygroscopic}. We define $G_{c,I}$ as the critical energy release rate for the interface between the core domain $V_{s,1}$ and shell domain $V_{s,2}$. Hereafter, we use the numbers 1 and 2 to respectively represent the variables in core and shell domains. The assumption of a sharp interface, $\Sigma$, would complicate the numerical modelling of chemo-mechanical fracture. Moreover, experiments have demonstrated a gradual chemical composition transition across the core-shell interface \cite{wu2016aligned}. To address this, we employ the phase field order parameter to smoothly interpolate the fracture energy across the core-shell interface, enabling a more accurate representation of debonding behaviour. An interface indicator $\zeta$ transitions from 0 (in the bulk material, far from the interface) to 1 (at the sharp interface), with the width of this diffuse zone controlled by a length scale parameter $\ell_{\zeta}$:
\begin{equation}
    \zeta-\ell_{\zeta}^2\nabla^2\zeta=0 \quad \text { in }  V_s,
\end{equation}
The boundary conditions are 
\begin{align}
& \zeta = 1 \quad \text { at }  \Sigma,\\
& \boldsymbol{\nabla} \zeta \cdot \mathbf{n}=0 \quad \text { at } \partial V_s,
\end{align}
And finally, we use the function $\left(1-\zeta\right)^2$ to interpolate the critical energy release rate across the diffuse zone:
\begin{equation}
    G_c=\left(1-\zeta\right)^2\left(G_{c,i}-G_{c,I}\right)+G_{c,I},
\end{equation}
with $i=1,2$ representing the core and shell, respectively. The value of $G_{c,I}$ depends on the approach to bond the two materials and can be measured by mechanical experiments such as a peeling test. For core-shell structured cathode materials, interface bonding is achieved via coating techniques, including: co-precipitation coating, dry coating, sol-gel coating and chemical vapor deposition (CVD). Unfortunately, to the authors' best knowledge, no data on the mechanical properties of the core-shell interface is available in the literature. Assumptions will be made for the value of $G_{c,I}$ based on the bulk material properties. \\

\begin{figure}[H]
    \centering
    \includegraphics[width=1\textwidth]{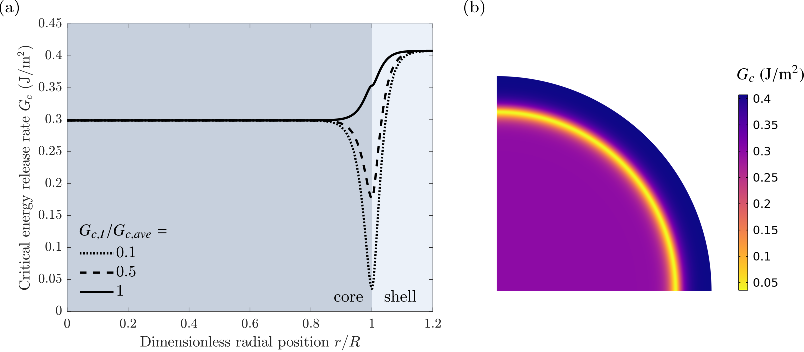}
    \caption{A diffuse representation of the core-shell interface: (a) Radial distribution of $G_c$ under 3 assumptions for $G_{c,I}$; (b) Visual representation of the variation of $G_c$ with the assumption $G_{c,I}=0.1G_{c,ave}$. }
    \label{fig:gcr}
\end{figure}

The radial distribution of $G_c$ within a spherical core-shell particle is given in
Fig. \ref{fig:gcr}(a) for 3 scenarios: (1) good bonding, where $G_{c,I}$ equals the average value of the bulk materials $G_{c,ave} = (G_{c,1} + G_{c,2}) / 2$, indicating an intimate contact coating; (2) weak bonding, where $G_{c,I}$ is reduced to half of $G_{c,ave}$; and (3) very weak bonding, where $G_{c,I}$ is only 10\% of $G_{c,ave}$. A contour map of $G_c$ for scenario (3) is displayed in Fig. \ref{fig:gcr}(b), suggesting that the core-shell interface would be more prone to cracking than bulk materials. Only a quarter section of the spherical particle is shown due to symmetry. Sun \textit{et al.} demonstrated through experiments that core-shell interface debonding, caused by structural mismatch and difference in volume changes between the core and shell, can lead to a sudden capacity drop \cite{sun2006novel,sun2006synthesis}. The influence of interface bonding strength on cracking behaviour will be analysed in Section \ref{sec:Results}. \\

In some modelling studies that address both bulk material fracture and debonding, a phase field fracture model is used for the bulk materials, while a cohesive zone model is applied to predict debonding \cite{tan2022phase,paggi2018fracture,quintanas2020phase}. By defining a diffuse zone for the core-shell interface, we can simultaneously solve both issues as cracking phenomena within the same framework. This approach requires only a single global phase field variable, 
$\phi$, to be solved across the entire domain.

\subsection{Diffusion hindered by mechanical damage}
\label{subsec:coupling}
Appropriate coupling of mechanical damage with electrochemical behaviour is far from straightforward. For solid-state lithium-ion batteries, cracking of electrode particles influences effective lithium-ion transport properties due to the dynamic nature of the electrode-electrolyte interface \cite{ruess2020influence}. The consequences of cracking in lithium-ion batteries with liquid electrolytes are more complex. Here, mechanical damage can cause internal cracks within the active material and external cracks that potentially expose new surfaces to the electrolyte. External cracks without electrolyte infiltration impede lithium-ion diffusion due to the loss of contact within the active material. The complexity of this process is compounded by factors such as the wettability of new surfaces, which influences the electrolyte infiltration behaviour \cite{sauter2020understanding,gao2020modeling}. If infiltration does occur, although it increases the active surface area, it can also promote transition metal (TM) dissolution and side reactions between the newly exposed material and the electrolyte \cite{friedrich2019capacity,liu2017intergranular}. Additionally, the low electrical conductivity of cathode materials such as lithium nickel manganese cobalt oxide (NMC) hinders electron flow along cracked surfaces that are no longer connected to the conductive matrix, thereby negatively impacting the overall electrochemical kinetics of a battery \cite{han2024computational}. \\

Some researchers have made initial attempts to incorporate these behaviours into simulations. Han \textit{et al.} \cite{han2024computational} assumed instant electrolyte infiltration into new cracks and the corrosion of the newly exposed surface, where a penalty factor was applied to the interfacial charge transfer kinetics. Allen \textit{et al.} \cite{allen2024using} proposed a similar approach, considering full electrolyte infiltration and a penalty factor scaling the exchange current density on the fractured surface. However, the values of these penalty factors are hypothetical, as experimental investigations of the electrochemical phenomena on new cracks are challenging. \\

For the Ni-rich layered materials examined in this work, cracks exposed to the electrolyte may trigger parasitic reactions, promoting the formation of spinel-like and rock-salt phases that hinder diffusion \cite{zhang2020problems}. Additionally, internal cracks obstruct diffusion due to loss of contact. Therefore, to account for the impact of mechanical damage on lithium-ion transport within the solid, we apply the degradation function $g(\phi)$ to the diffusion coefficient, expressed as $D=D_0g(\phi)$, where $D_0$ is the original diffusion coefficient of lithium ions in the active material.

\subsection{Boundary conditions for intercalation}
\label{subsec:bc}
The numerical experiments utilise a constant current-constant voltage (CC-CV) profile for the intercalation process. The CV stage is terminated when the average current density drops to the cut-off value, which is 10\% of the current density during the CC stage. We assume a continuous chemichal potential at the core-shell interface \cite{Wu2017}, which establishes the following relationship between the concentration in the two materials at the core-shell interface \cite{tu2024influence}:
\begin{equation}
    J_1(r=R)=J_2(r=R),
    \label{eq:J1equalJ2}
\end{equation}
\begin{equation}
    \mu_1(r=R)=\mu_2(r=R),
\end{equation}
\begin{equation}
    r=R:\quad c_1=U_{r e f, 1}^{-1}\left[\frac{\Omega_{2} \sigma_{h, 2}-\Omega_{1} \sigma_{h, 1}}{F}+U_{r e f, 2}\left(c_{2}\right)\right],
    \label{eq:c1}
\end{equation}

\noindent Here, $r$ is the radial position within a spherical particle. $\mathrm{U_{r e f, 1}}$ and $\mathrm{U_{r e f, 2}}$ are the open circuit potential of the core and shell. $R$ denotes the radius of the core and $h$ is the thickness of the shell. The other boundary and initial conditions are described as
\begin{equation}
    r=0:\quad \frac{\partial c_1}{\partial r}=0,
\end{equation}
\begin{equation}
    t=0:\quad c_1=c_{1,0}\quad \mathrm{and}\quad c_2=c_{2,0},
\end{equation}
\begin{equation}
    r=R+h:\quad J_2(r=R+h)=J_0\quad \mathrm{during\ the\ CC\ stage},
\end{equation}

The state of lithiation (SOL) of the particle is defined as
\begin{equation}
    \mathrm{SOL}=\frac{\int_{V_s}c\ \mathrm{d}V}{V_1c_{max,1}+V_2c_{max,2}},
\end{equation}
where \(c_{max,i}\ (i=1,2)\) are the maximum concentration of lithium in the core and shell materials. $\mathrm{SOL}=0\%$ and $100\%$ respectively represent fully delithiated and fully lithiated states.\\

The flux $J_0$ during CC stage depends on the C-rate $C$ as,
\begin{equation}
    J_0=\frac{V_1c_{max,1}+V_2c_{max,2}}{\mathrm{surface\ area}}\frac{C}{3600\ \mathrm{s}}
\end{equation}

\section{Results and discussion}
\label{sec:Results}
The coupled model described above is numerically implemented using the finite element method in COMSOL Multiphysics. 
We investigate a system consisting of a high-nickel core with a lower-nickel-content shell, a configuration that has shown promise as a high-performance cathode \cite{ma2019comparative,tan2020recent,lu2014modified}.
Specifically, a spherical core-shell particle of $\mathrm{LiNi_{0.8}Mn_{0.1}Co_{0.1}O_{2}}$ (NMC811) coated with $\mathrm{LiNi_{0.5}Mn_{0.3}Co_{0.2}O_{2}}$ (NMC532) is analysed. \\
\begin{table}[H]
\centering
\begin{tabular}{cccc}
\toprule  
Parameter&Symbol&Core (NMC811)&Shell (NMC532)\\
\midrule 

Maximum lithium concentration&$c_{max}$&51765 $\mathrm{mol/m^3}$ \cite{Chen2020}&49000 $\mathrm{mol/m^3}$ \cite{sharma2022asynchronous}\\
Partial molar volume of lithium&$\Omega$&$\mathrm{7.88\times 10^{-7}\ m^3/mol}$ \cite{Biasi2017}&$\mathrm{4.86\times 10^{-7}\ m^3/mol}$ \cite{Biasi2017}\\
Diffusion coefficient&$D$&$\mathrm{3.26\times 10^{-14}\ m^2/s}$ \cite{tu2024influence}&$\mathrm{2.48\times 10^{-14}\ m^2/s}$ \cite{su2015enhancing}\\
Young’s modulus&$E$&$\mathrm{230\ GPa}$ \cite{Sharma2023} &$\mathrm{201\ GPa}$ \cite{Sharma2023}\\
Poisson’s ratio&$\nu$&$\mathrm{0.253}$ \cite{Sharma2023}&$\mathrm{0.253}$ \cite{Sharma2023}\\
Toughness & $G_c$&$0.299\ \mathrm{N/m}$&$0.408\ \mathrm{N/m}$ \cite{Sharma2023}\\
Length scale & $\ell$ &0.23 µm&0.27 µm\\
\bottomrule 
\end{tabular}   
\caption{Material properties of the NMC core-shell structure.}
\label{table:2}
\end{table}
The material properties used in the numerical model are listed in Table \ref{table:2}. 
Given the inherent complexity of the coupled model, constant material properties are used for simplicity, despite evidence from previous studies showing their dependency on lithium-ion concentration \cite{tu2024influence,Chen2020}.
We assume the length scale for the diffuse interface to be $\ell_{\zeta}=0.1$ µm, which represents a transition zone that is much smaller than the shell thickness. The lithiation process starts with 10\% of the maximum concentration in the shell and the corresponding value in the core, determined by the boundary condition described in Section \ref{subsec:bc}. \\

Due to the challenges associated with conducting fracture tests on cathode materials, we determine the phase field input parameters from indentation tests performed on NMC secondary particles. Sharma \textit{et al.} \cite{Sharma2023} carried out indentation tests on NMC811 and NMC532 and inferred fracture toughnesses of $K_c=0.271\ \mathrm{MPa\ m^{0.5}}$ and $K_c=0.296\ \mathrm{MPa\ m^{0.5}}$, respectively. The value of $G_c$ can then be estimated as $G_c=(1-\nu)^2K_c^2/E$ \cite{irwin1957analysis}. The length scale is intrinsically defined by the choices of tensile strength ($\sigma_c$) and toughness ($G_c$), with the material characteristic length scale defined as $\ell_{ch}=(K_c / \sigma_c)^2$ and the phase field length scale given by $\ell=27 \ell_{ch} / 256$ \cite{Mandal2019}, for the so-called AT model. The tensile strength of NMC materials is reported to be $\sigma_c=184 \mathrm{MPa}$ \cite{Wheatcroft2023}.\\

To reduce computational cost, the simulations are performed using an axisymmetric geometry, with the contour results shown through 2D cross-sections. \\

\begin{figure}[H]
    \centering
    \includegraphics[width=1\textwidth]{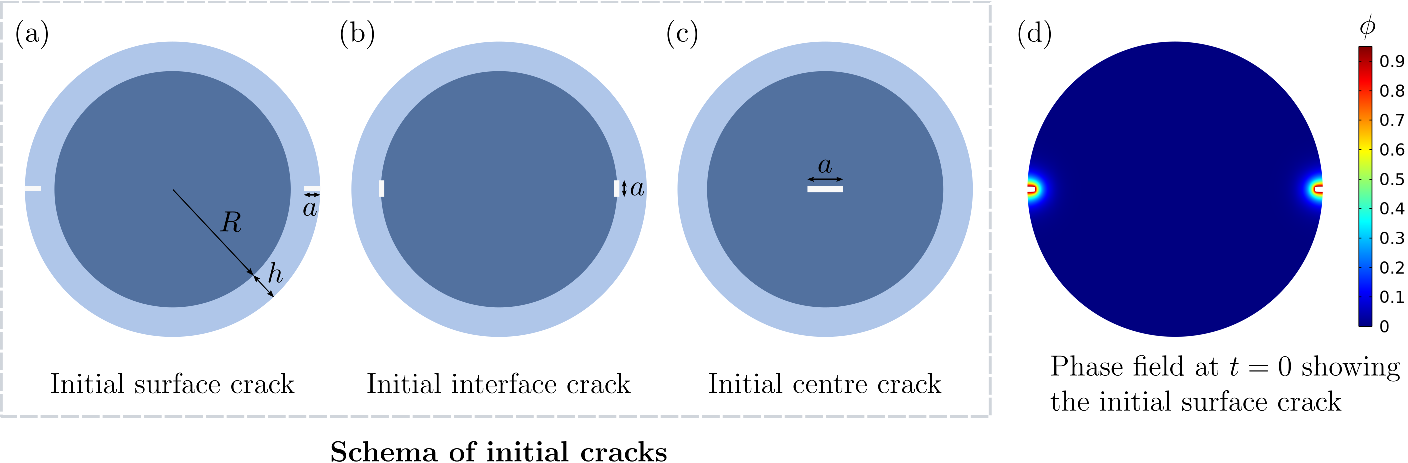}
    \caption{Key definitions and boundary value problem. (a)-(c) Schematics of three types of initial cracks in spherical core-shell particles; and (d) contour plot showing that a diffuse initial crack is used instead of a geometric one. In all figures, fully cracked regions ($\phi > 0.95$) are removed for the sake of better crack visualisation.}
    \label{fig:initialcrack}
\end{figure}

Experiments have revealed that cathode particles may have pre-existing cracks caused by mechanical stresses during calendering, winding, or synthesis \cite{Parks2023,beuse2021,Heenan2020}. These defects can initiate particle cracking. In our numerical experiments, three types of initial cracks are investigated: those located on the particle surface, at the core-shell interface, and at the centre of the core, as illustrated in Figs. \ref{fig:initialcrack}(a)-(c). 
Diffuse initial cracks are employed, as depicted in Fig. \ref{fig:initialcrack}(d) and, to facilitate crack visualisation, cracked regions ($\phi>0.95$) are removed from the contour plots. Initial cracks are introduced numerically by assigning a non-zero history field within the region corresponding to the initial crack length at $t=0$, expressed as:
\begin{equation}
    \mathcal{H}=\alpha_0\ \mathrm{exp} \left(-100\frac{z^2}{\ell^2} \right),
\end{equation}
where $\alpha_0=10^{12}\ \mathrm{J/m^3}$ and $z$ denotes the distance to the initial crack plane.\\

\begin{figure}[H]
    \centering
    \includegraphics[width=0.6\textwidth]{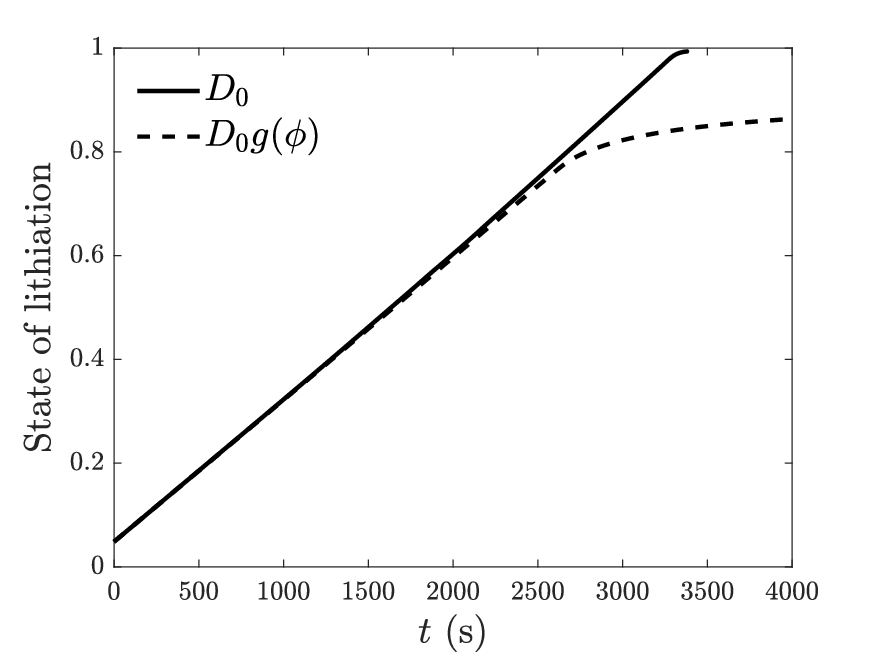}
    \caption{Evolution of the state of lithiation (SOL) for a cracked particle. When assuming that the mechanical damage impedes diffusion ($D=D_0g(\phi)$), the state of lithiation evolution shows a capacity fade compared to full lithiation without coupling between damage and diffusion ($D_0$).}
    \label{fig:dg}
\end{figure}

The effect of design and operating parameters on cracking behaviours is investigated, focusing on the core-shell system dimensions, C-rate, and core-shell bonding strength. The geometric parameters, illustrated in Fig. \ref{fig:initialcrack}, include core size $R$, shell thickness $h$, and initial crack size $a$. The severity of cracking and capacity fade are quantified. We define the normalised crack volume, $\bar{a}_c$, as the ratio between the fully cracked ($\phi>0.95$) volume to the total volume of the particle:\\
\begin{equation}
    \bar{a}_c=\frac{\int_{V_s} H(\phi-0.95) \mathrm{d} V_s}{V_s} ,
\end{equation}
where $V_s$ represents total volume of the particle and  $H$ is the Heaviside step function. Capacity degradation is assessed by observing the final state of lithiation at the cut-off current. Fig. \ref{fig:dg} shows the state of lithiation evolution of a NMC811@NMC532 particle with an initial surface crack during lithiation, with and without considering the degradation of diffusion. Both models predict cracking propagation; however, the model with $D_0$ reaches full lithiation, while the model used in this study, which considers the interplay between mechanical damage and diffusion $D=D_0g(\phi)$, indicates a capacity loss.

\subsection{Validation with experimental observation}
\label{subsec:validation}

\begin{figure}[H]
    \centering
    \includegraphics[width=1\textwidth]{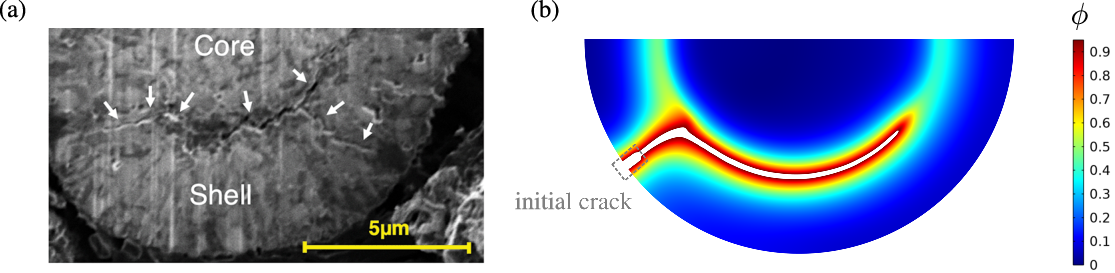}
    \caption{Validation against experimental observations: (a) SEM image of a core-shell particle showing shell fracture and core-shell debonding after cycling, reproduced from Ref. \cite{Brandt2020}; (b) Simulated phase field results, obtained using the same inputs as the experiment, show good visual agreement with the experimental observations.}
    \label{fig:validation}
\end{figure}

Before conducting a predictive parametric analysis, model predictions are validated against experimental findings.
Quantitatively characterising particle cracking during electrochemical cycling remains challenging. 
Electron microscopy (EM) and X-ray imaging techniques have been used to study cracks in cathode particles caused by fabrication or operation \cite{Parks2023,Heenan2020}.
However, cracking characterisation in cycled core–shell particles remains largely unexplored. Brandt et al. \cite{Brandt2020} employed scanning electron microscopy (SEM) to examine a cycled NMC core–shell particle and observed both shell fracture and core–shell debonding after a single C/3 half-cycle, as shown in Fig. \ref{fig:validation}(a). A comparison is performed with this experimental result to validate the computational model described in Section \ref{sec:Methods}. In their study, Brandt et al. \cite{Brandt2020} investigated a particle consisting of an NMC811 core and a concentration-gradient shell with NMC532 at the surface. NMC811 and NMC532 are also used in our numerical case studies, with their material properties listed in Table \ref{table:2}. To represent the concentration-gradient shell, a linear radial interpolation of material properties is applied between NMC811 and NMC532.
As no SEM image of the particle prior to cycling is available, an initial surface crack is assumed to originate from the fabrication process. Fig. \ref{fig:validation}(b) shows the simulated phase field after lithiation at C/3, capturing crack propagation within the shell and core–shell debonding, which closely resembles the cracking pattern observed in the SEM image in Fig. \ref{fig:validation}(a).

In the following sections, studies with three types of initial cracks are conducted individually. Representative cracking patterns and their impact on capacity loss are analysed in detail using the example of an initial surface crack in Section \ref{subsec:shell}. Similar behaviours are observed for initial interface cracks, which are examined in Section \ref{subsec:core-shell}. Finally, the case with an initial centre crack, which exhibits distinct cracking patterns, is discussed in Section \ref{subsec:core}.\\

\subsection{Cracking with an initial crack on the shell surface}

\label{subsec:shell}
\begin{figure}[H]
    \centering
    \includegraphics[width=0.9\textwidth]{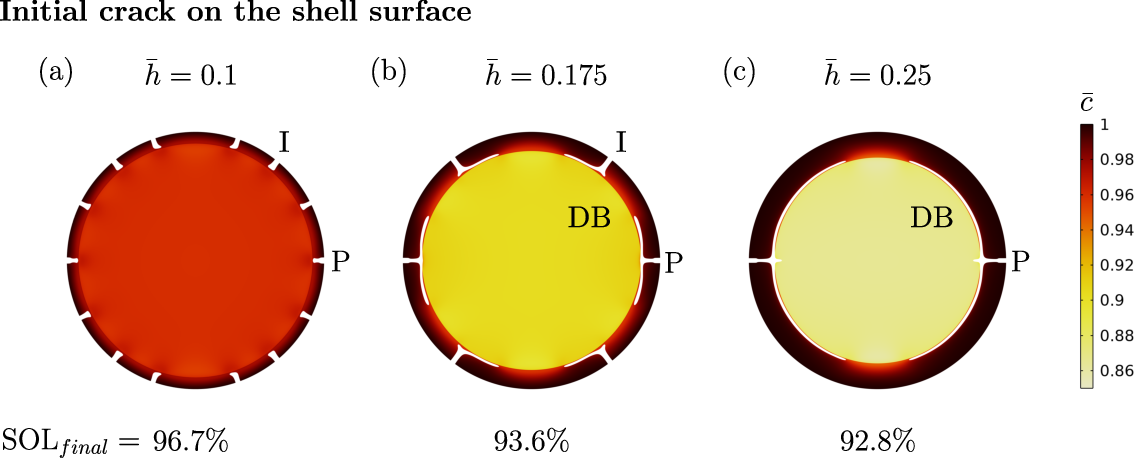}
    \caption{Three cracking phenomena are identified: propagation (P) of the initial crack in the shell, initiation (I) of new surface cracks, and core-shell debonding (DB). Three representative cracking patterns, combining these phenomena, are observed for an initial surface crack: (a) P+I; (b) P+I+DB; and (c) P+DB.}
    \label{fig:scrack}
\end{figure}

Initial surface cracks can result in various cracking patterns, depending on the design and operating parameters. 
The representative final cracking patterns are illustrated in Fig. \ref{fig:scrack}, where fully fractured regions ($\phi > 0.95$) are removed for better crack visualisation. Three distinct cracking phenomena are identified: propagation (P) of the initial crack in the shell, initiation (I) of new surface cracks, and core-shell debonding (DB). The cracking patterns are combinations of these three phenomena. It is observed that the initial crack propagates (P) in the shell in all cracking patterns, driven by particle swelling. New surface crack initiation (I) and core-shell debonding (DB) may occur individually, when a single cracking mechanism dominates, or simultaneously. The final concentration field $c$ and accessible SOL are also shown in Fig. \ref{fig:scrack}. A key observation is that interfacial debonding (see Figs. \ref{fig:scrack}(b)(c)) impedes lithium-ion transport into the core, thereby resulting in a lower final accessible SOL compared to Fig. \ref{fig:scrack}(a). \\

\begin{table}[H]
\centering
\begin{tabular}{p{3cm}p{3cm}p{3cm}p{5cm}}
\toprule  
Parameters&P+I&P+I+DB&P+DB\\
\midrule 
$\bar{h}$&0.1, 0.125&0.15, 0.175&\textbf{0.2}, 0.225, 0.25, 0.275, 0.3\\
$R\ \text{(µm)}$&&2, 3&\textbf{4}, 5, 6, 7, 8\\
$C$&&0.5&\textbf{1}, 2, 3\\
$\bar{a}$&&&0.2, \textbf{0.3}, 0.4, 0.5\\
$G_c^i/G_c^{ave}$&&&0.1, 0.5, \textbf{1}\\
\bottomrule 
\end{tabular}
\caption{Final cracking patterns obtained for the case of an initial surface crack and every choice of parameters. The reference values are marked in bold. For most case studies, a failure model involving propagation (P) and core-shell debonding (DB) is observed.}
\label{table:shell}
\end{table}

The studied parameters and the resulting final cracking patterns for various parameter values are presented in Table \ref{table:shell}. The reference values are marked in bold. When analysing one parameter, the others remain fixed, taking their reference values. The parameters include design parameters of the core-shell system (core radius $R$ and relative shell thickness $\bar{h} = h/R$; see Fig. \ref{fig:initialcrack} for the geometric schema), an operating parameter (C-rate $C$), and parameters influenced by the fabrication process (relative defect size $\bar{a} = a/h$ and interfacial bonding strength $G_c^i / G_c^{ave}$).
As shown in Table \ref{table:shell},
all relative initial crack sizes ($\bar{a}$=0.2, 0.3, 0.4, and 0.5) result in the same cracking pattern. Additionally, the evolution of normalised crack volume ($\bar{a}_c$) over time is similar across all initial crack sizes. Therefore, the effects of the initial crack size will not be discussed in detail hereafter. The key cracking mechanisms will be analysed in Section \ref{subsubsec:pattern}, followed by the influence of various parameters investigated in Sections \ref{subsubsec:bonding} and \ref{subsubsec:rhc}.

\subsubsection{Cracking mechanisms of new surface cracks and interfacial debonding}
\label{subsubsec:pattern}
\begin{figure}[H]
    \centering
    \includegraphics[width=0.7\textwidth]{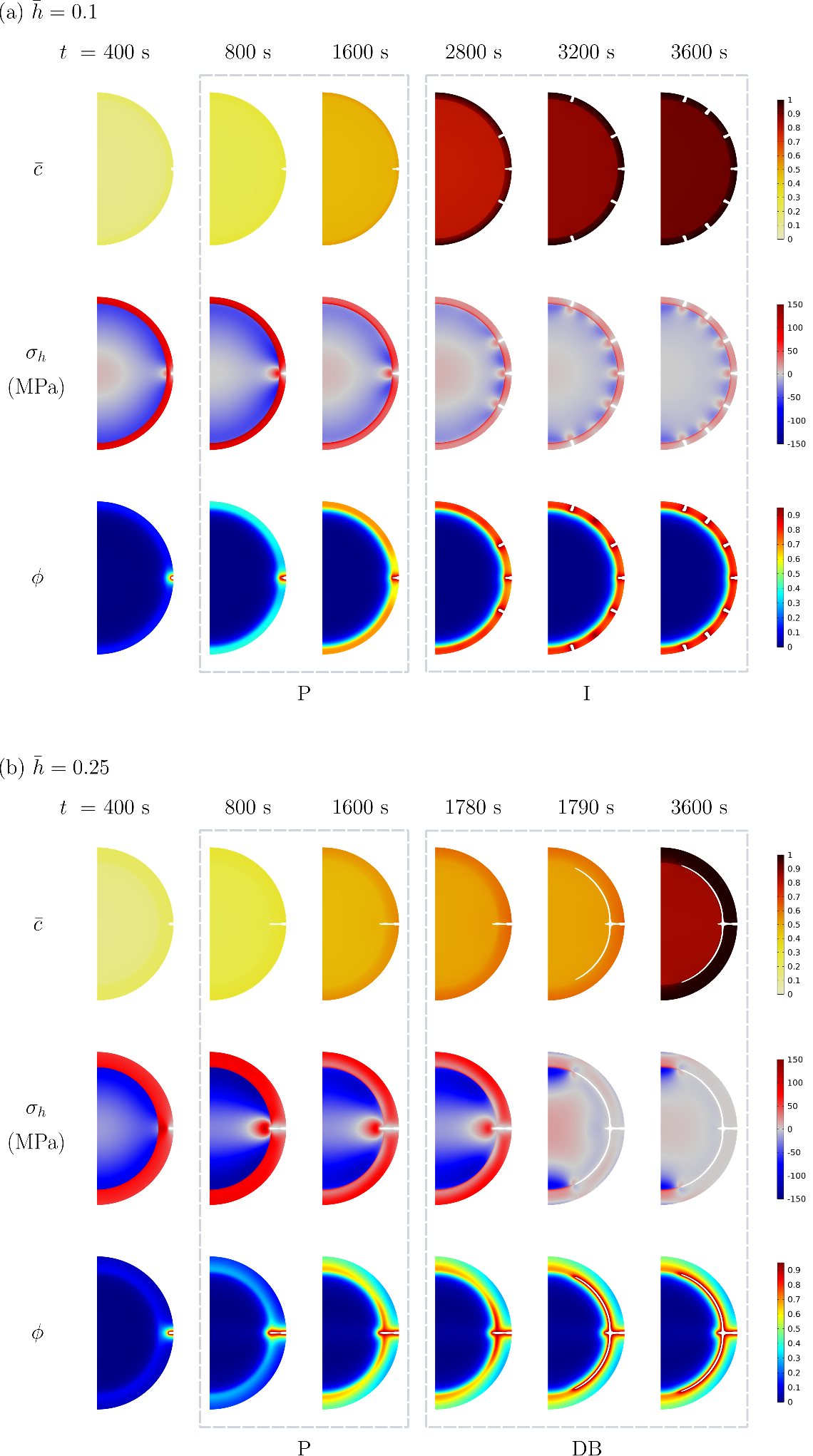}
    \caption{Evolution of the dimensionless concentration $\bar{c}$, hydrostatic stress $\sigma_h$, and phase field $\phi$, displayed for two shell thicknesses: (a) results for a thin shell ($\bar{h}$=0.1) showing initial crack propagation (P) followed by new crack initiation (I); (b) results for a relatively thick shell ($\bar{h}$=0.25) depicting initial crack propagation (P) followed by core-shell debonding (DB). Only half of the particle section is shown due to symmetry.}
    \label{fig:I}
\end{figure}

To understand the cracking mechanisms of initiation (I) of surface cracks and interfacial debonding (DB), the evolution of the dimensionless concentration $\bar{c}$, hydrostatic stress $\sigma_h$, and phase field $\phi$ is shown in detail for exemplar thin-shell and thick-shell particles in Fig. \ref{fig:I}. For both cases, the initial crack propagates (P) first in the shell, followed by new crack initiation for the thin shell or interfacial debonding for the thick shell.\\

For the particle with a thin shell ($\bar{h}=0.1$, see Fig.~\ref{fig:I}(a)), a substantially larger fraction of lithium is accommodated in the core compared with the shell. As a result, the core undergoes pronounced expansion, constrained by the less expansive shell. This difference in volume change arises from the greater lithium accumulation in the core and the relatively smaller partial molar volume of the shell. The resulting strain mismatch generates tensile hoop stresses in the shell, with the maximum located at the core–shell interface.This high stress drives the formation of new cracks that penetrate the shell at multiple locations, as shown in Fig.~\ref{fig:I}(a).
When damage occurs, the stress is significantly diminished as the degradation function $g(\phi)$ reduces the effective stiffness of the material. These surface cracks have little impact on the concentration field, since they do not substantially obstruct diffusion, resulting in a relatively high final SOL of 96.7\%, which is higher than the cases where debonding occurs. \\

When the shell is thicker, it accommodates more lithium ions and undergoes greater outward expansion, while still constraining the expansion of the core. As in the case with a thin shell, the mismatch in volume change generates large hoop stresses in the shell, with the maximum being at the core–shell interface. The increased thickness results in a higher difference in stress levels across the thickness, which localises damage at the interface, triggering interfacial debonding, as shown in Fig. \ref{fig:I}(c) for a particle with $\bar{h}=0.25$. Stress decreases once damage occurs. Interfacial debonding hinders lithium-ion transport from the shell to the core, reducing the final accessible SOL to 92.8\% for $\bar{h}=0.25$.

\subsubsection{Effects of core-shell interface bonding}
\label{subsubsec:bonding}

\begin{figure}[H]
    \centering
    \includegraphics[width=0.6\textwidth]{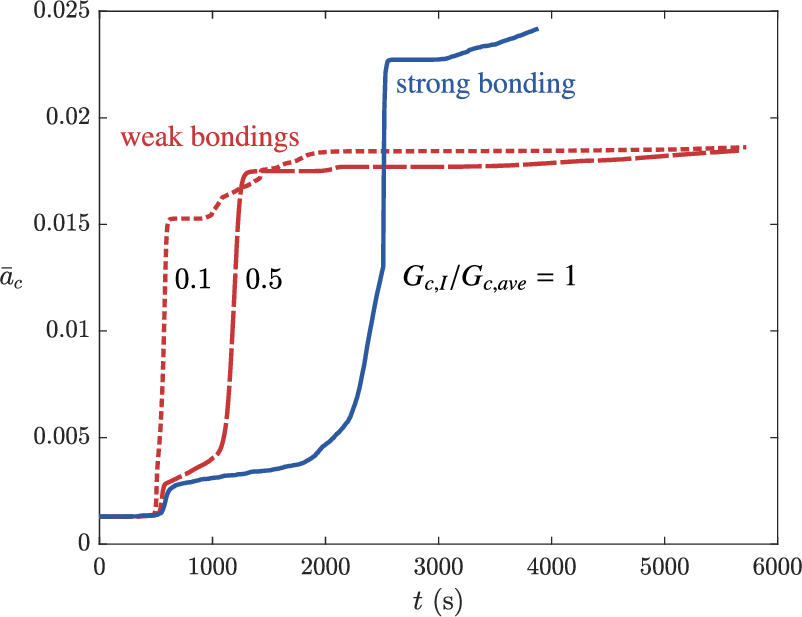}
    \caption{Evolution of the normalised crack volume ($\bar{a}_c$) over time under three assumptions for the interface bonding: weak bondings ($G_{c,I}=0.1G_{c,ave}$ and $0.5G_{c,ave}$) and strong bonding ($G_{c,I}=G_{c,ave}$). The sudden rise of $\bar{a}_c$ indicates debonding, which happens earlier for weaker bondings.}
    \label{fig:gc}
\end{figure}

The evolution of the normalised crack volume $\bar{a}_c$ for three different interface bonding strengths is depicted in Fig. \ref{fig:gc}: both weak ($G_{c,I}=0.1G_{c,ave}$ and $0.5G_{c,ave}$) and good ($G_{c,I}=G_{c,ave}$) bonding scenarios are considered. They all lead to the same cracking pattern: propagation of the initial crack in the shell, followed by core-shell debonding. Under the same loading condition (intercalation), core-shell debonding occurs significantly earlier for weak bondings, as indicated by the abrupt increase in $\bar{a}_c$ shown in Fig. \ref{fig:gc}. Once debonding initiates, lithium diffusion into the core is abruptly blocked in the debonded region, suppressing further core expansion and thus limiting interfacial debonding. Over time, however, the shell continues to take up and diffuse lithium ions to the core through intact regions, leading to gradual expansion. This renewed expansion can trigger limited additional debonding, which appears as a slight upward shift in the $\bar{a}_c$ curve.
The final accessible SOL is 74.8\%, 82.8\%, and 92.2\% for $G_{c,I}/G_{c,ave}=0.1$, 0.5 and 1, respectively. Early debonding obstructs lithium-ion diffusion into the core, reducing the accessible capacity. These results align with the experimental findings of a sudden capacity drop caused by core-shell debonding \cite{sun2006novel,sun2006synthesis}.

Strengthening the core-shell bond is crucial to prevent early-stage debonding. Wu \textit{et al.} \cite{wu2016aligned} reported that an intimate core-shell contact can be achieved through ball-milling and temperature control below 100 °C. Transmission electron microscopy (TEM) and etching X-ray photoelectron spectroscopy (XPS) have conﬁrmed a gradual chemical transition zone between core and shell materials, suggesting a smooth $G_c$ transition where $G_{c,I}=G_{c,ave}$. In the interest of optimising the microstructural design to enhance the chemo-mechanical behaviours of the particle, subsequent calculations assume an optimal interfacial design with $G_{c,I}=G_{c,ave}$.

\subsubsection{Effects of core radius, relative shell thickness, and C-rate}
\label{subsubsec:rhc}

\begin{figure}[H]
    \centering
    \includegraphics[width=1\textwidth]{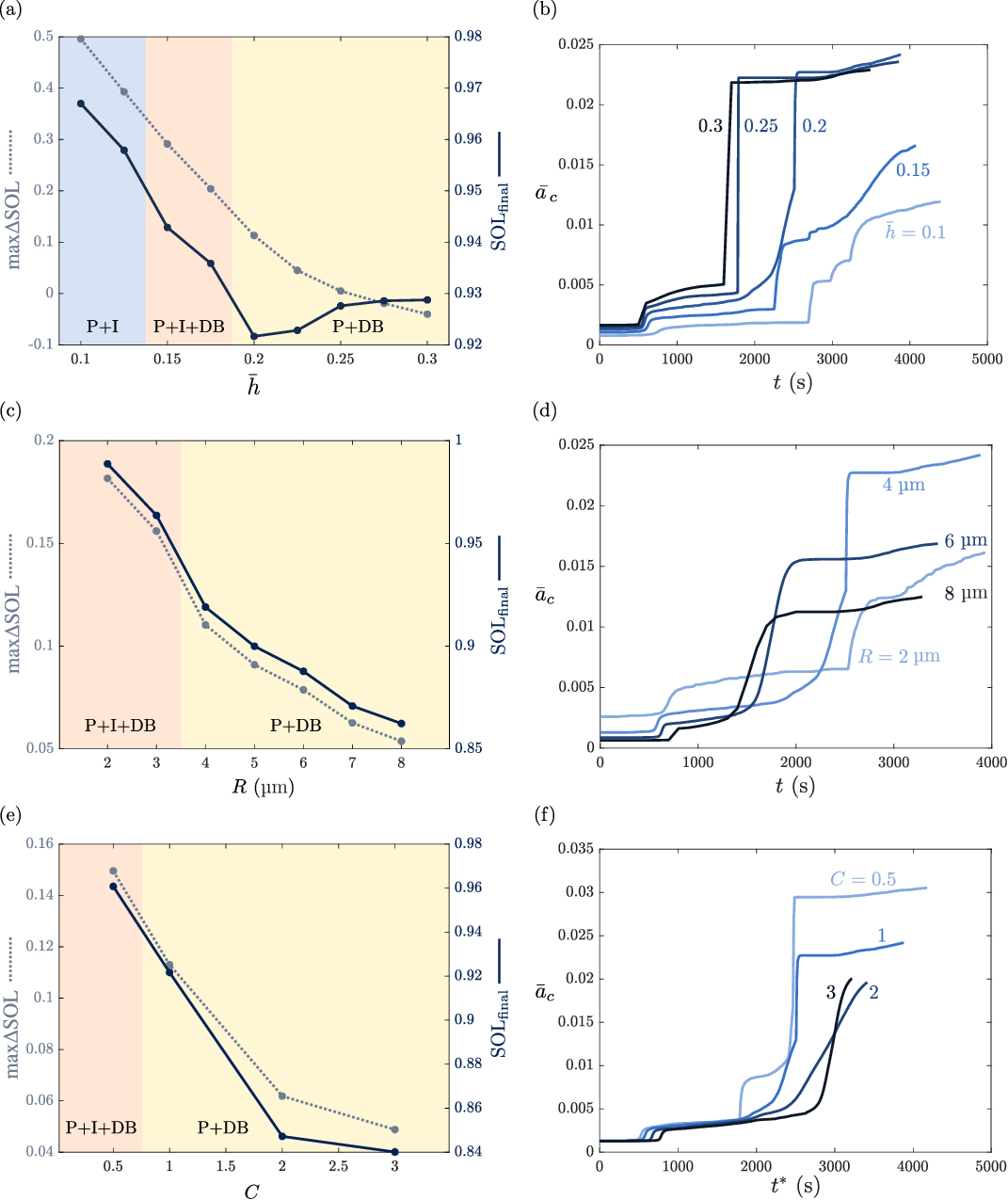}
    \caption{Influence of the parameters on cracking and capacity loss: maximum $\Delta\mathrm{SOL}$ during lithiation and final accessible SOL under varying (a) $\bar{h}$ , (c) $R$, and (e) $C$, corresponding to different cracking patterns: P+I, P+I+DB, or P+DB; and the evolution of $\bar{a}_c$ over time under different values of (b) $\bar{h}$, (d) $R$, and (f) $C$.}
    \label{fig:solcrack}
\end{figure}

As discussed in Section \ref{subsubsec:pattern}, the cracking pattern is influenced by the distribution of lithium ions in the core and shell, which dominates the volume change of the core and shell, thereby affecting the evolution of the stress field and the phase field $\phi$. When significantly more lithium ions are distributed in the core than in the shell, it leads to new surface cracks; while when the core and shell accommodate similar amounts of lithium ions, core-shell debonding occurs. In intermediate cases, surface crack initiation and debonding occur simultaneously. 
To quantify the difference of overall state of lithiation in the core and shell, we define: $\Delta\mathrm{SOL}=\mathrm{SOL_{core}}-\mathrm{SOL_{shell}}$. 
The value of $\Delta\mathrm{SOL}$ is influenced by the relative volumes of the core and shell, which are determined by $\bar{h}$, as well as by the concentration gradient which is affected by the core size $R$ and C-rate $C$.
The maximum $\Delta\mathrm{SOL}$ and final accessible SOL during lithiation under varying relative thickness $\bar{h}$, core radius $R$, and C-rate $C$, are illustrated in Figs. \ref{fig:solcrack}(a)(c)(e), respectively. 
When the shell is very thin ($\bar{h}=0.1,\ 0.125$), $\mathrm{max} \Delta\mathrm{SOL}$ exceeds 0.4, leading to initiation of surface cracks. When the shell thickens ($\bar{h}=$0.2 to 0.3), or the core radius increases beyond 4 µm, or $C$ reaches 2 or 3, core-shell debonding becomes dominant, reducing the final accessible SOL. \\
The normalised crack volume ($\bar{a}_c$), shown in Figs. \ref{fig:solcrack}(b)(d)(e), quantifies the evolution of crack initiation and propagation.
For thick shells ($\bar{h}$=0.2 to 0.3), interfacial debonding occurs rapidly, characterised by the steep increases in $\bar{a}_c$ in Fig. \ref{fig:solcrack}(b). The thicker the shell is, the earlier debonding takes place. Once rapid debonding happens, it impedes diffusion and significantly reduces the chemo-mechanical loading, causing a plateau in $\bar{a}_c$ in Figs \ref{fig:solcrack}(b), which can also be observed in Figs. \ref{fig:solcrack}(d)(f). After debonding, lithiation continues until the cut-off current is reached at the outer surface. The shell continues to accommodate and diffuse lithium ions to the core through the intact regions. For the thickest shell ($\bar{h}=0.3$), the shell is able to store slightly more lithium before reaching the cut-off condition, resulting in a modestly higher final capacity compared to $\bar{h}=0.2$, as shown in Fig. \ref{fig:solcrack}(a).
For $R=4,\ 6,\ 8$ µm, core-shell debonding is observed and can be characterised by the rise of $\bar{a}_c$ in Fig. \ref{fig:solcrack}(d). Since the crack width, governed by the length scale $\ell$, is relatively smaller for bigger particles, $\bar{a}_c$ is consequently smaller for larger particles exhibiting the same cracking pattern. Initiation of new surface cracks is characterised by small steps in $\bar{a}_c$, as observed for $\bar{h}=0.1,\ 0.15$, $R=2$ µm, and $C=0.5$. In Fig. \ref{fig:solcrack}(f), time is normalised in reference to $C=1$, where $t^*=t\cdot C$. For fast discharging rates of $2C$ and $3C$, interfacial debonding takes place rapidly, blocking the transport of lithium ions into the core. As a result, the lithium ions become saturated in the shell quickly, reaching cut-off condition during the CV stage much faster.\\

Overall, particle size and C-rate, which control the concentration gradient within the particle, have a greater impact on the final accessible SOL than the relative shell thickness. When the other parameters are fixed, all relative shell thickness from $\bar{h}=0.1$ to $0.3$ result in more than 92\% of final SOL. However, when $R$ exceeds 5 µm, or $C$ is 2 or higher, the final SOL drops below 90\% because of core-shell debonding.

\subsection{Cracking with an initial crack at the core-shell interface}
\label{subsec:core-shell}
\begin{figure}[H]
    \centering
    \includegraphics[width=1\textwidth]{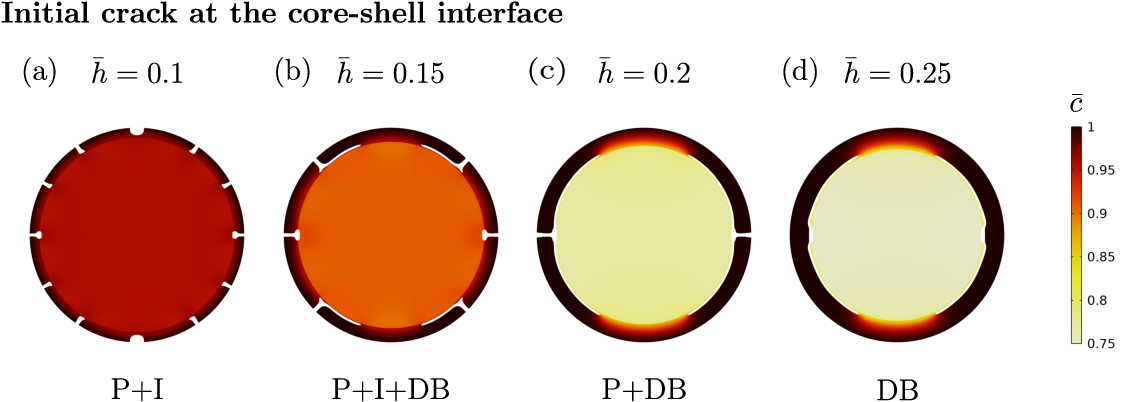}
    \caption{The same cracking phenomena are identified: propagation (P) of the initial crack in the shell, initiation (I) of new surface cracks, and core-shell debonding (DB). Four representative cracking patterns are observed for an initial surface crack: (a) P+I; (b) P+I+DB; and (c) P+DB; and a newly observed pattern, purely debonding (d) DB.}
    \label{fig:icrack}
\end{figure}

For the case with an initial crack at the core-shell interface, the range of parameters studied and the reference values of the parameters remain the same as in Section \ref{subsec:shell}, except for the choice $\bar{a} = a/h = 0.6$ instead of 0.3, due to symmetry. As for the analysis of initial surface cracks in Section \ref{subsec:shell}, the initial crack size  has a minimal effect on the cracking behaviours.
The same cracking phenomena are identified as before: propagation (P) of the initial crack in the shell, initiation (I) of new surface cracks, and core-shell debonding (DB). The same combinations of these phenomena are also observed. Additionally, a new cracking pattern is recognised: purely debonding, as depicted in Fig. \ref{fig:icrack}(d).\\

\begin{table}[H]
\centering
\begin{tabular}{p{3cm}p{2.5cm}p{2.5cm}p{2.5cm}p{2.5cm}}
\toprule  
Parameters&P+I&P+I+DB&P+DB&DB\\
\midrule 
$\bar{h}$&0.1&0.15&\textbf{0.2}&0.25, 0.3\\
$R\ \text{(µm)}$&&2&\textbf{4}&6, 8\\
$C$&&0.5&\textbf{1}, 2, 3&\\
$G_c^i/G_c^{ave}$&&&0.5, \textbf{1}&0.1\\

\bottomrule 
\end{tabular}
\caption{Final cracking patterns obtained for the case of an initial interface crack and every choice of parameters. The reference values are marked in bold.}
\label{table:interface}
\end{table}

\begin{figure}[H]
    \centering
    \includegraphics[width=0.8\textwidth]{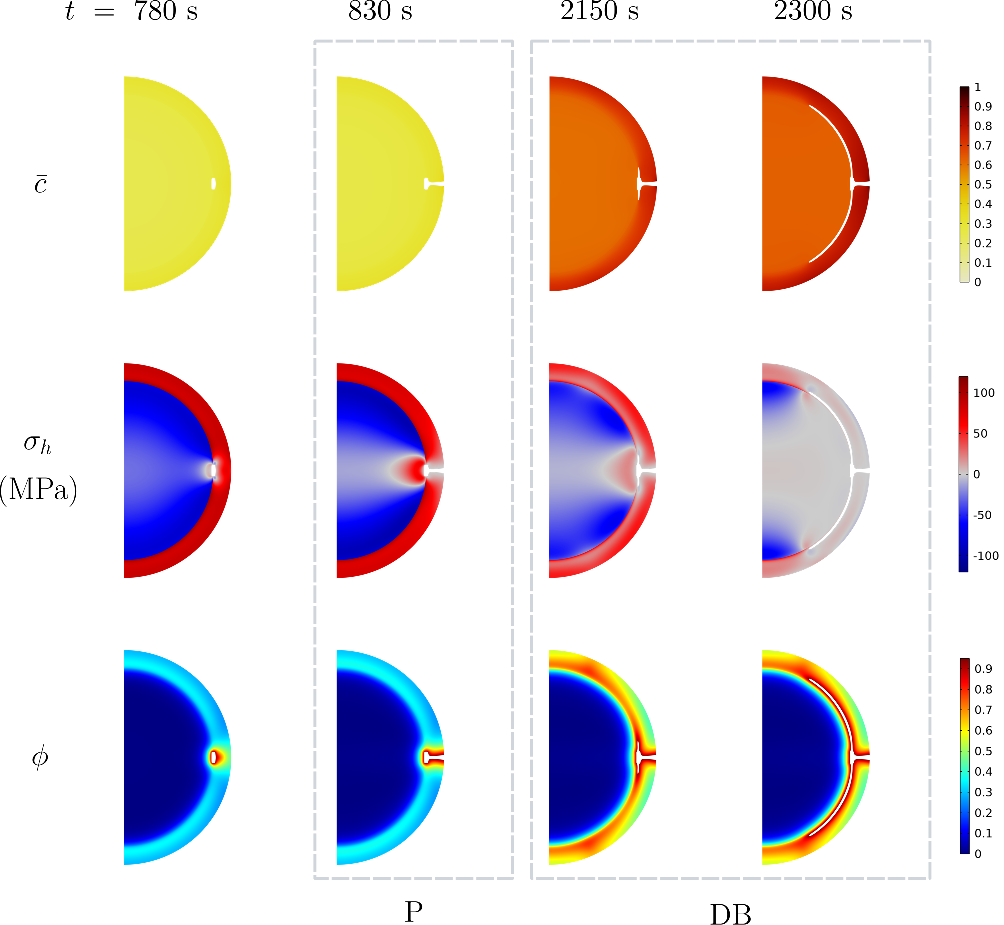}
    \caption{Evolution of $\bar{c}$, $\sigma_h$, and $\phi$ within a particle with an initial interface crack, illustrating the crack propagation (P) through the shell and subsequently at the core-shell interface (DB). Only half of the particle section is shown due to symmetry.}
    \label{fig:PDB}
\end{figure}

The studied parameters and the resulting final cracking patterns for various parameter values are listed in Table \ref{table:interface}. The reference values are indicated in bold.
A particle with the reference values as parameters undergoes propagation within the shell, followed by interfacial debonding. The evolution of $\bar{c}$, $\sigma_h$, and $\phi$ are illustrated in Fig. \ref{fig:PDB}. At the beginning of the lithiation process, the concentration field exhibits a nearly radial symmetry. The presence of the initial crack at the interface introduces heterogeneity in the stress and damage fields, leading to the opening of a crack through the shell. The subsequent cracking bahaviour closely resembles that observed for an initial surface crack, where high stress in the shell near the interface drives debonding.
The effects of $\bar{h}$, $R$, and $C$ on cracking are also consistent with those in the cases with an initial surface crack. Notably, a new finding is the transition to purely interfacial debonding under specific conditions, such as weak interface bonding ($G_{c,I} = 0.1G_{c,ave}$) or greater hoop stress induced by thicker shells ($\bar{h} = 0.25$, $0.3$) and larger particles ($R = 6$, $8$ µm).

\subsection{Cracking with an initial crack in the centre of the core}
\label{subsec:core}

\begin{figure}[H]
    \centering
    \includegraphics[width=1\textwidth]{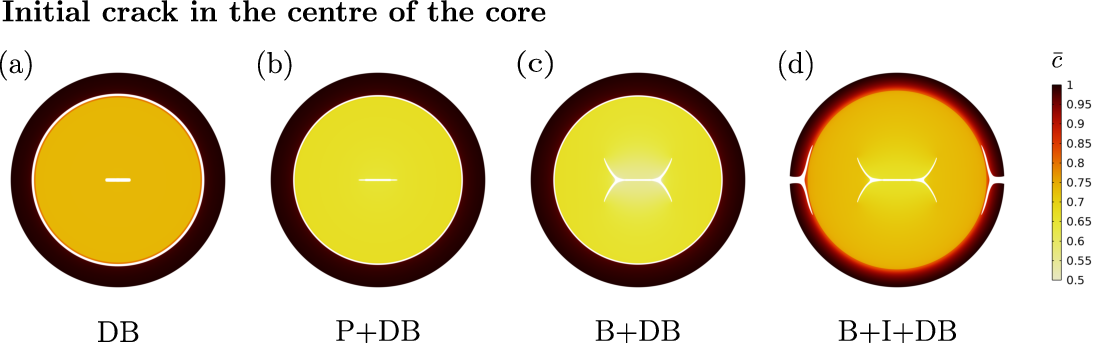}
    \caption{The observed cracking phenomena include radial propagation (P) of the initial crack, initiation (I) of surface cracks, core-shell debonding (DB), and branching (B). Four representative cracking patterns are identified for an initial central crack: (a) DB, (b) P+DB, (c) B+DB, and (d) B+I+DB.}
    \label{fig:ccrack}
\end{figure}
\begin{figure}[H]
    \centering
    \includegraphics[width=0.8\textwidth]{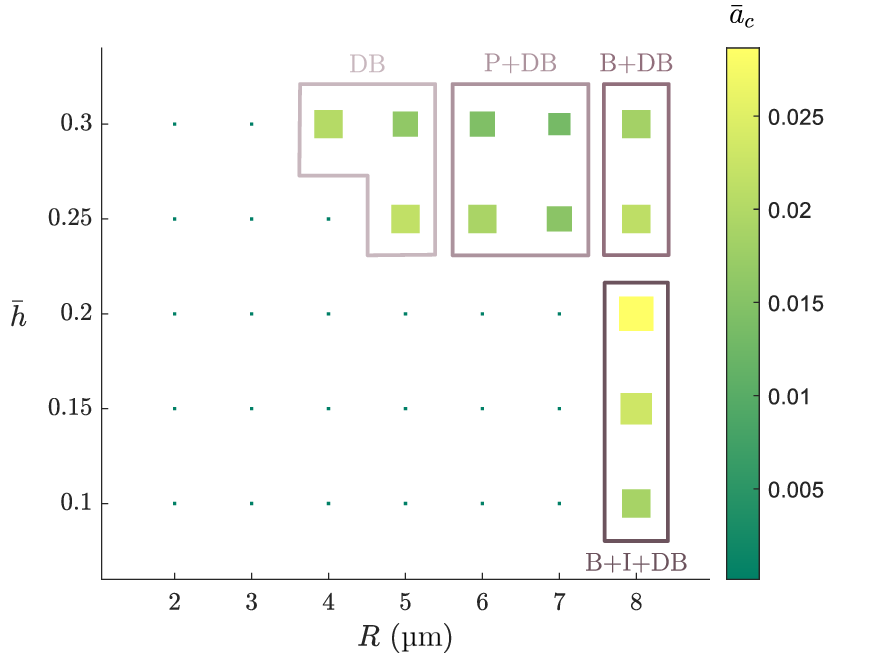}
    \caption{Visual representation of the final normalised crack volume, $\bar{a}_c$, as a function of $R$ and $\bar{h}$ in the presence of an initial central crack. The dots indicate no cracking. The boxes highlight the corresponding cracking patterns.}
    \label{fig:ccrack2c}
\end{figure}

The final case study considered involves an initial crack located at the centre of the core. Despite the presence of a crack at the centre, the core and shell expand in an almost spherically symmetric manner. Compared to surface and interface cracks, the internal central crack leads to lower local fluctuation in concentration and generates smaller stresses. 
When examining the same combinations of parameters as in Sections \ref{subsec:shell} and \ref{subsec:core-shell}, almost no crack propagation or initiation is observed. Therefore, we further investigate cases with a higher C-rate as reference ($C=2$), while varying $R$ from 2 to 8 µm and $\bar{h}$ from 0.1 to 0.3, with $\bar{a}=a/R=0.6$ and $G_{c,I}/G_{c,ave}=1$. \\
The final cracking patterns are shown in Fig. \ref{fig:ccrack2c}. Initiation (I) of surface cracks and core-shell debonding (DB) can be identified, as in previous Sections \ref{subsec:shell} and \ref{subsec:core-shell}.
The central initial crack can lead to either propagation (P) along the same radial direction or a newly observed phenomenon: branching (B), as illustrated in Fig. \ref{fig:ccrack}. 
Branching is when the crack tip bifurcates, leading to the formation of two or more crack paths. Crack branching can be understood as a consequence of excess energy available at the crack tip, which cannot be fully dissipated through a single crack propagation event \cite{bleyer2017dynamic}. It is typically associated with high crack tip velocities \cite{karma2004unsteady,henry2013fractographic}.\\

\begin{figure}[H]
    \centering
    \includegraphics[width=0.7\textwidth]{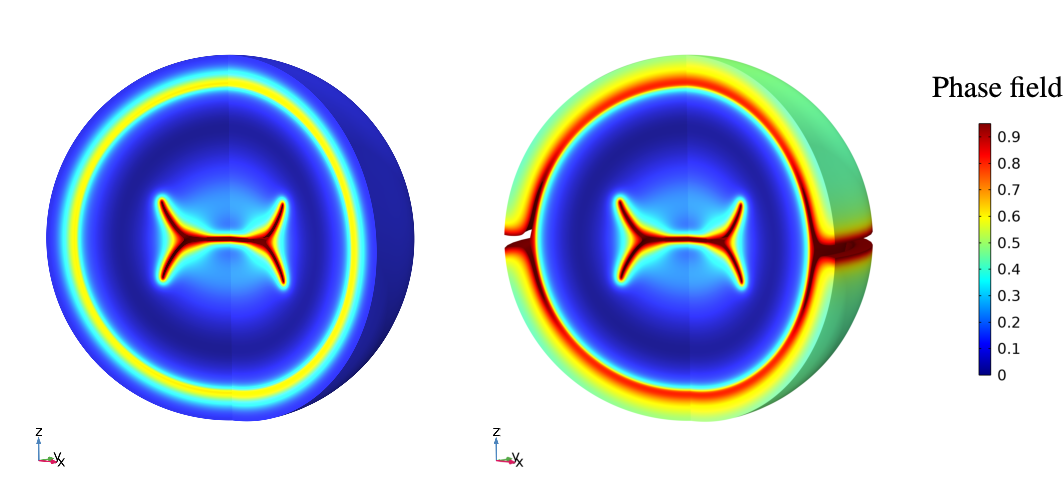}
    \caption{Evolution of the phase field in a particle with $R=8$ µm, $\bar{h}=0.2$ undergoing branching, shown in 3D. A portion of the particle is removed to reveal the internal structure.}
    \label{fig:3dbranching}
\end{figure}

The influence of the geometric parameters $R$ and $\bar{h}$ on cracking behaviour is illustrated in Fig. \ref{fig:ccrack2c}. The final normalised crack volume, $\bar{a}_c$, quantifies the severity of cracking, with the boxes representing groups of various cracking patterns. The dots indicate cases where $\bar{a}_c$ is nearly zero, corresponding to parameter combinations that result in no cracking.
Particles with $R\leq7$ µm and $\bar{h}\leq0.2$ experience no cracking. Larger particles or thicker shells generate higher stresses at the core-shell interface, leading to interfacial debonding. Very large particles ($R=8$ µm) undergo branching, as illustrated in Fig. \ref{fig:3dbranching}. Similarly, Klinsmann \textit{et al.} \cite{klinsmann2016modeling} predicted branching of spherical cathode particles when $R\geq10$ µm during lithiation under various C-rates. The branching of large particles can be attributed to increased crack tip velocity, driven by larger regions of high tensile stress that accelerate the crack tip. A smaller radius and a thinner shell should be chosen to mitigate mechanical degradation in this scenario.

\section{Conclusions}                                                                                             
\label{sec:conclusions}
We present a novel chemo-mechano-damage framework that predicts mechanical degradation and the resulting capacity fade in electrode particles of lithium-ion batteries. By coupling the electrochemical behaviour with fracture mechanics, our model uniquely connects particle cracking to capacity loss which is a crucial link often overlooked.
Using the phase field method, our model addresses both bulk material fracture and interfacial debonding simultaneously. 
Through numerical case studies of NMC core-shell particles, we investigate how design and operating parameters influence cracking, providing insights for improving battery durability.
The main findings are 

\begin{itemize}
  \item The location of initial defects plays a dominant role. Under the same conditions, initial surface and interface cracks are more detrimental than those at the centre due to higher stresses in the shell. In large particles (core radius $R\geq8$ µm), initial central cracks may branch and split the particle due to accelerated crack tip in larger tensile regions.
  
  \item Concentration distribution governs cracking pattern. If significantly more lithium ions are concentrated in the core (when the shell is very thin or the concentration gradient is small), high hoop stresses develop throughout the shell, potentially leading to new surface cracks. In contrast, if the lithium-ion distribution is more balanced between core and shell, the shell tends to expand more outward, inducing high hoop stresses near the core-shell interface that lead to interfacial debonding. Notably, interfacial debonding hinders lithium transport in the radial direction, making it more detrimental to capacity.
  
  \item Particle size and C-rate dominate capacity loss. The final accessible state of lithiation (SOL) is predominantly influenced by particle size and C-rate, which govern the concentration gradient. In the case of an initial surface crack, when $R$ exceeds 5 µm or $C$ reaches 2 or higher, the final SOL drops below 90\% due to core-shell debonding.
\end{itemize}

The findings also provide guidance for designing core-shell particles that minimise mechanical degradation and result in improved battery performance. For example, the results suggest that one should optimise the manufacturing process to reduce initial imperfections in the shell. In addition, the core-shell interface has been shown to be a weak link; therefore, the use of advanced synthesis techniques \cite{wu2016aligned} or graded materials \cite{Sun2009,sun2012nanostructured,ju2014optimization,xu2019progressive} presents suitable avenues.\\ 

Further experimental and computational studies are needed to explore phenomena related to particle cracking, such as electrolyte infiltration and its consequences. Moreover, future research should account for the polycrystalline structure to gain a more comprehensive understanding of particle cracking. In addition, experiments have shown that mechanical properties deteriorate during cycling \cite{jangde2025mechanical,xu2017mechanical}. Therefore, further studies addressing the long-term performance of cathode particles could incorporate fatigue effects to capture ageing behaviours \cite{Ai2022}.

\section*{Acknowledgements}

B. Wu was supported by the EPSRC Faraday Institution Multi-Scale Modelling project (EP/S003053/1, grant number FIRG003). E. Mart\'{\i}nez-Pa\~neda acknowledges financial support from UKRI's Future Leaders Fellowship programme [grant MR/V024124/1] and the National Research Foundation of Korea (NRF) through the MSIT grant RS-2024-00397400.

\end{document}